\documentclass[prl,showpacs,superscriptaddress,twocolumn,longbibliography]{revtex4-1}

\usepackage[colorlinks=false]{hyperref}
\usepackage{color}
\usepackage[usenames,dvipsnames]{xcolor}
\usepackage{amsmath,amsthm,amssymb}
\usepackage{graphicx}
\usepackage{float}
\usepackage{epsfig}
\usepackage{bm}
\usepackage{mathrsfs}
\usepackage{multirow}
\usepackage[all]{xy}
\usepackage{pbox}
\usepackage{verbatim}
\usepackage{braket}
\usepackage{mathtools}
\usepackage{bm}
\usepackage{tikz}
\usepackage{xcolor}
 
\usepackage{mathtools}

\usepackage{mathtools}
\usepackage{enumerate}   

\DeclareMathOperator{\Tr}{Tr}

\newcommand{\er}[1]{Eq.~\eqref{#1}}

\newcommand{\beq}{\begin{equation}}
\newcommand{\eeq}{\end{equation}}

\newcommand{\WW}{\mathbb W}
\newcommand{\JJ}{\mathbb J}
\newcommand{\RR}{\mathbb R}
\newcommand{\LL}{\mathbb L}
\newcommand{\R}{\mathcal{R}}
\newcommand{\K}{\mathcal{K}}
\newcommand{\A}{\mathcal{A}}
\newcommand{\modL}[1]{#1\;{\rm mod}\;L}
\newcommand{\Str}{S_{\rm Tri}}
\newcommand{\Sntr}{S_{\rm Top}}

\begin{document}  

\title{Topological phases in the dynamics of the simple exclusion process}

\author{Juan P. Garrahan}
\affiliation{School of Physics and Astronomy, University of Nottingham, Nottingham, NG7 2RD, UK}
\affiliation{Centre for the Mathematics and Theoretical Physics of Quantum Non-Equilibrium Systems,
University of Nottingham, Nottingham, NG7 2RD, UK}
\author{Frank Pollmann}
\affiliation{Department of Physics and Institute for Advanced Study,
Technical University of Munich, 85748 Garching, Germany}
\affiliation{Munich Center for Quantum Science and Technology (MCQST), Schellingstr.\ 4, D-80799 M\"unchen}

\begin{abstract}
    We study the dynamical large deviations of the classical stochastic symmetric simple exclusion process (SSEP) by means of numerical matrix product states. We show that for half-filling, long-time trajectories with a large enough imbalance between the number hops in even and odd bonds of the lattice belong to distinct symmetry protected topological (SPT) phases. Using tensor network techniques, we obtain the large deviation (LD) phase diagram in terms of counting fields conjugate to the dynamical activity and the total hop imbalance. We show the existence of high activity trivial and non-trivial SPT phases (classified according to string-order parameters) separated by either a critical phase or a critical point. 
    Using the leading eigenstate of the tilted generator, obtained from infinite-system density matrix renormalisation group (DMRG) simulations, we construct a near-optimal dynamics for sampling the LDs, and show that the SPT phases manifest at the level of rare stochastic trajectories.  We also show how to extend these results to other filling fractions, and discuss generalizations to asymmetric SEPs. 
\end{abstract}

\maketitle

\noindent 
{\bf\em Introduction.--} 
Certain problems in classical stochastic dynamics bear close resemblance at the technical level to problems in quantum many-body. One such is computing distributions of dynamical observables (see e.g. Refs.~\cite{Lecomte2007,Garrahan2009, Touchette2009, Esposito2009, Garrahan2018, Jack2020, Limmer2021}). In the long time limit, the statistics of a time-extensive function of a stochastic trajectory (such as the dynamical activity \cite{Garrahan2007, Maes2020} or a time-integrated current \cite{Derrida2007}) often obeys a large deviation (LD) principle, whereby its distribution and moment generating function (MGF) scale exponentially in time \cite{Lecomte2007,Garrahan2009, Touchette2009, Esposito2009, Garrahan2018, Jack2020, Limmer2021}. In the LD regime, all relevant information is contained in the functions in the exponent, known as the rate function for the probability and the scaled cumulant generating function (SCGF) for the MGF, with rate function and SCGF related by a Legendre transform \cite{Lecomte2007,Garrahan2009, Touchette2009, Esposito2009, Garrahan2018, Jack2020, Limmer2021}. 
This is the generalization of the ensemble method of statistical mechanics to dynamics \cite{Eckmann1985, Ruelle2004, Merolle2005}, with trajectories being the microstates, the long-time limit the thermodynamic limit, the MGF the partition sum, and rate function and SCGF the entropy density and free energy density, respectively.

\begin{figure}[ht!!!]
    \centering
    \includegraphics[width=\columnwidth]{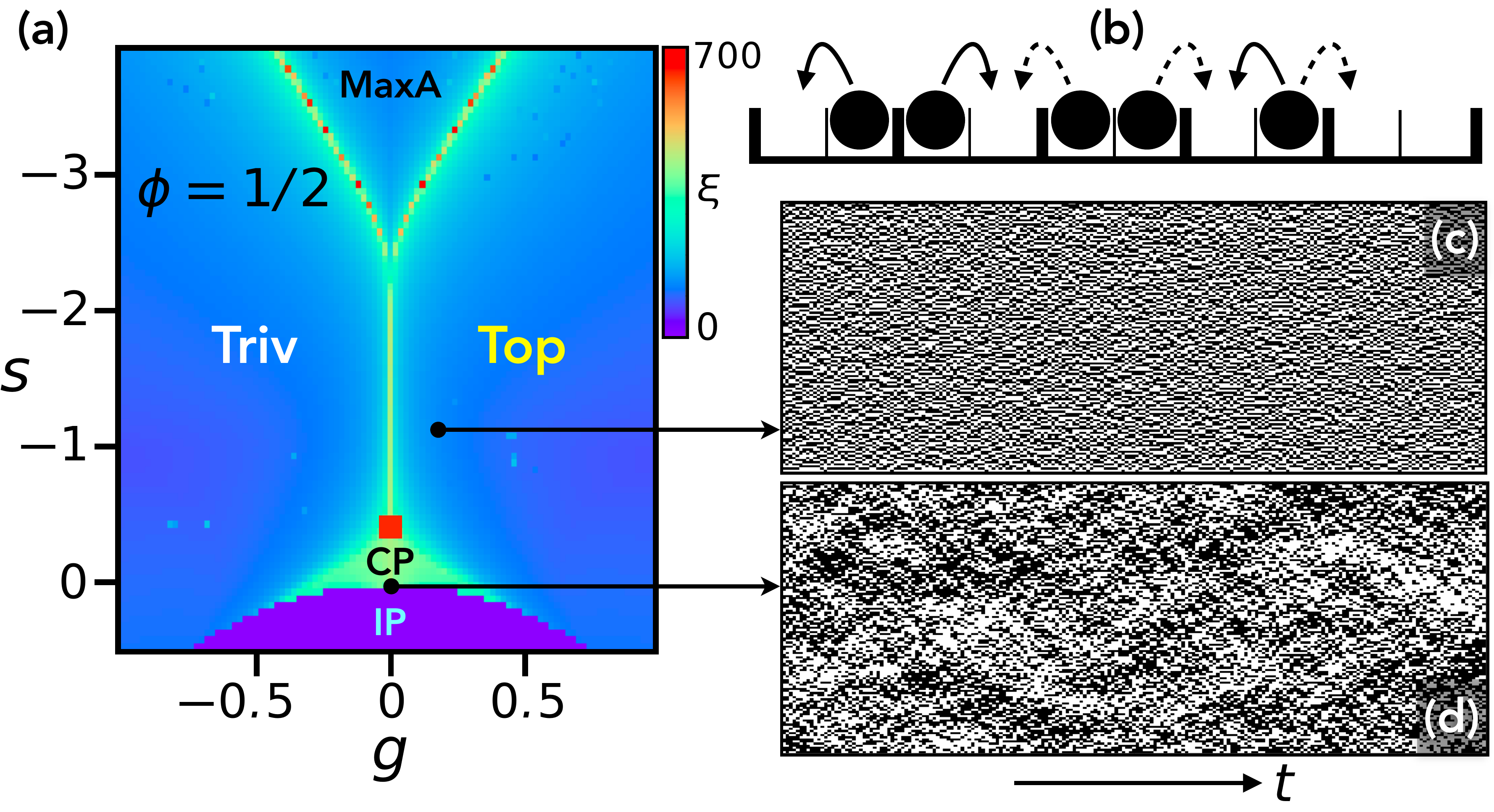}
    \caption{(Colour online)
        (a) LD phase diagram of the SSEP at half-filling, $\phi=1/2$, as function of counting fields $g$ (for staggered number of jumps $\K^{(2)}$) and $s$ (for time-integrated escape rate $\R$). The two distinct symmetric phases are denoted by ``Triv'' (for trivial) and ``Top'' (for topological). For $s < s_{L}=1/\sqrt{2}-1$ (indicated by the red square) the transition between SPT phases is continuous, while for $s \geq s_{L}$ they are separated by a critical phase (light green, CP). For $s < -2$ there is a phase of maximal activity with antiferromagnetic (AF) order (MaxA). The symmetric phases and the AF phase are separated by a line of Ising critical points. At $s>0$ there is an inactive phase (purple, IP). (Shown is the correlation length $\xi$ from infinite-system DMRG simulations with bond dimension $\chi=64$.) 
        (b) Observable $\K^{(2)}$: jumps across odd bonds are counted as $+1$ while jumps across even bonds are counted $-1$. (c) Rare trajectory from the ``Top'' phase ($g=0.2, s=1.1$) sampled with (approximately) optimal dynamics (see main text), $N=128, t_{\rm max}=10^4$. (d) Typical trajectory at $\phi = 1/2$ for comparison. 
         } 
    \label{fig1}
\end{figure}

In particular, the SCGF can be obtained \cite{Lecomte2007,Garrahan2009, Touchette2009, Esposito2009, Garrahan2018, Jack2020, Limmer2021}
as the largest eigenvalue of a deformation, or {\em tilting}, of the Markov generator of the dynamics
\footnote{
    We focus on continuous-time Markov chains for simplicity, but similar ideas apply to discrete Markov chains and to diffusions.
}.
The problem of calculating dynamical LDs is thus equivalent to finding the ground state of a stoquastic Hamiltonian \cite{Bravyi2006}. Furthermore, away from the LD regime of long times, the classical problem becomes equivalent to calculating a quantum partition sum (with particular time boundaries) \cite{Causer2022}. These analogies have allowed to obtain precise analytical results for LD behavior from certain one-dimensional systems from know exact properties of associated quantum spin chains
\cite{Appert-Rolland2008,Karevski2017}, and have also motivated the use of numerical tensor network methods to accurately estimate LDs functions in kinetically constrained systems \cite{Banuls2019, Helms2019, Helms2020,Causer2020, Causer2021, Causer2022}. 

In this paper, we expand on the analogies above by showing that in the one-dimensional simple exclusion process (SEP) ---a paradigmatic model for the study of non-equilibrium stochastic dynamics (for reviews see \cite{Blythe2007,Mallick2015}) --- 
large non-homogeneous fluctuations in the dynamics can belong to distinct symmetry protected topologically (SPT) phases \cite{GuTEFspt2009, Pollmann2009aSPTone, Chen2011Classonedimensionalspinsystems}. 
These SPT phases are characterized by topological invariants that require the presence of an unbroken symmetry. 
In particular, the defining property of one-dimensional SPT phases is how particular bulk symmetries act anomalously on the edge.
The most prominent example is the Haldane phase, realized by the gapped spin-1 Heisenberg chain with its spin-$1/2$ edge modes: the bulk is symmetric with respect to $SO(3)$ whereas the edges transform projectively under $SU(2)$ \cite{Haldane83a,Haldane83b}. 
A direct consequence of the anomalous action of the symmetry on the edges are modes at zero energy.
While SPT phases do not have any local order parameters, non-local (string) order parameters can be derived to detect them \cite{denNijs89, Pollmann2012Detection}. 

For simplicity, we consider the case of symmetric hopping rates, or SSEP  (but comment on the applicability to the asymmetric SEP, or ASEP, towards the end). The LDs of the SEP are well studied in terms of fluctuations of the dynamical activity, i.e., the number of hops in a trajectory, and time-integrated particle current \cite{Bodineau2007, Appert-Rolland2008, Lecomte2012, Jack2015, Karevski2017}. For the activity, studies have revealed the existence of distinct dynamical phases (away from typical diffusive dynamics), specifically an inactive and clustered phase, separated by first-order phase transition from a critical phase of a high activity and ``hyperuniform'' phase. In the language of the ground state of the quantum XXZ spin chain \cite{Sachdev2011}, these correspond to the ferromagnetic phase and the Luttinger liquid phase, respectively. We show here that when tilting is with respect to the number of hops with a stagger set by the filling fraction, dynamical SPT phases emerge, and the associated rare trajectories can be sampled efficiently from the numerical matrix-product state (MPS) solution of the tilted generator (see Fig.~1).

\begin{figure}[t!!!]
    \centering
    \includegraphics[width=\columnwidth]{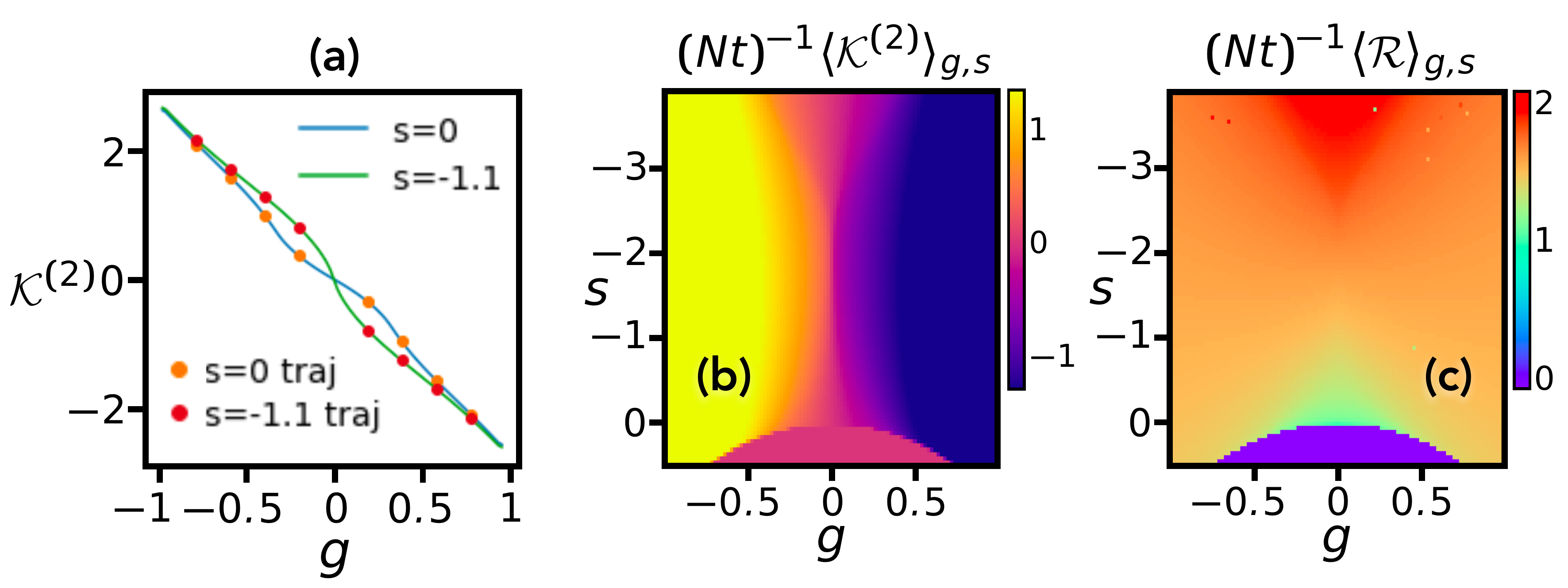}
    \caption{
        (Colour online) (a) Average staggered number of jumps $\lim_{t,N \to \infty}\langle \K^{(2)} \rangle_{g,s}/(N t)$ as a function of $g$ for $s=0$ (blue) and $s=-1.1$ (green) for half-filling $\phi=1/2$ (from iMPS with $\chi=128$). 
        Each symbol is the corresponding value from a single sampled trajectory ($N=128$, $t=10^3$) at the same conditions. (b) Phase diagram in terms of average $\K^{(2)}$ (per unit time and length). (c) Phase diagram in terms of average time-integrated escape rate $\R$.
        } 
    \label{fig2}
\end{figure}

\smallskip

\noindent 
{\bf\em Model and dynamical large deviations.--} 
The one-dimensional SSEP \cite{Blythe2007, Mallick2015} is a system of particles on a lattice with excluded volume interactions. We denote a configuration by $x = n_{1:N}$, with $n_{j}=0, 1$ indicating an empty or occupied site, respectively.  The master
equation for the evolution of the probability vector $\ket{P_t} = \sum_{x} P_t(x)\ket{x}$ (with $\{ \ket{x} \}$ the configuration basis) is $\partial_t \ket{P_t} = \mathbb{W} \ket{P_t}$. 
Particles can hop only to empty neighboring sites, with the same rate (which we set to unity) for left or right jumps in the SSEP. The Markov generator reads
\begin{align}
    \WW &= 
    \frac{1}{2} \sum_j \left( X_j X_{j+1} + Y_j Y_{j+1} + Z_j Z_{j+1} - 1 \right) ,
    \label{W}
\end{align}
where $X_j = \sigma_j^+ + \sigma_j^-$, $Y_j = -i (\sigma_j^+ + \sigma_j^-)$ and $Z_j = \sigma_j^+ \sigma_j^-$, are operators with Pauli matrices acting non-trivially on site $j$. The XY terms generate the nearest neighbor hopping, $\JJ_j = \frac{1}{2} (X_j X_{j+1} + Y_j Y_{j+1}) = \sigma_j^+ \sigma_{j+1}^- + \sigma_j^- \sigma_{j+1}^+$, while $\RR =  \sum_j \left( 1 - Z_j Z_{j+1} \right)$ is the escape rate operator (so that $\WW = \JJ - \RR$ with 
$\JJ = \sum_j \JJ_j$). The generator above is (minus) the Hamiltonian of the spin-$1/2$ ferromagnetic XXZ quantum spin chain at the stochastic (or Heisenberg) point \cite{Sachdev2011}. 
The continuous-time Markov dynamics defined by \er{W} is realized in terms of stochastic trajectories, $x_{0:t} = (x_0 \to x_{t_1} \to \cdots \to x_t)$, with $t_1 \cdots t_K$ the times when transitions occur. Dynamical observables are time-extensive functions of trajectories, $\A(x_{0:t})$. From the probability $\pi(x_{0:t})$ of realising $x_{0:t}$ in the dynamics we can obtain the distribution of a dynamical observable, $P_t(\A) = \sum_{x_{0:t}} \pi(x_{0:t}) \delta[\A-\A(x_{0:t})]$, and its moment generating function, $Z_t(s) = \sum_{x_{0:t}} \pi(x_{0:t}) e^{-s \A(x_{0:t})}$. For long times, these obey an LD principle, $P_t(\A) \asymp e^{-t \varphi(\A/t)}$ and $Z_t(s) \asymp e^{t \theta(s)}$, with $\varphi(a)$ and $\theta(s)$ the rate function and SCGF, respectively \cite{Lecomte2007,Garrahan2009, Touchette2009, Esposito2009, Garrahan2018, Jack2020, Limmer2021}.

We consider the joint LDs of two dynamical observables. The first one is the {\em time-integrated escape rate}, $\R(x_{0:t}) = \int_t \langle x_t | \RR | x_t \rangle$, which provides the same information as the dynamical activity 
\cite{Garrahan2009}. The second, is the {\em difference in the activity between odd and even bonds} of the lattice: if $\K_j(x_{0:t})$ denotes the total number of particle hops in a trajectory between sites $j$ and $j+1$, we define $\K^{(2)} = \sum_j (-)^{j+1} K_j$. The superscript indicates that this dynamical observable has spatial period two and is the appropriate one for half-filling, $\phi = 1/2$ (we generalize for other fillings below).

\begin{figure}[t!!!]
    \centering
    \includegraphics[width=\columnwidth]{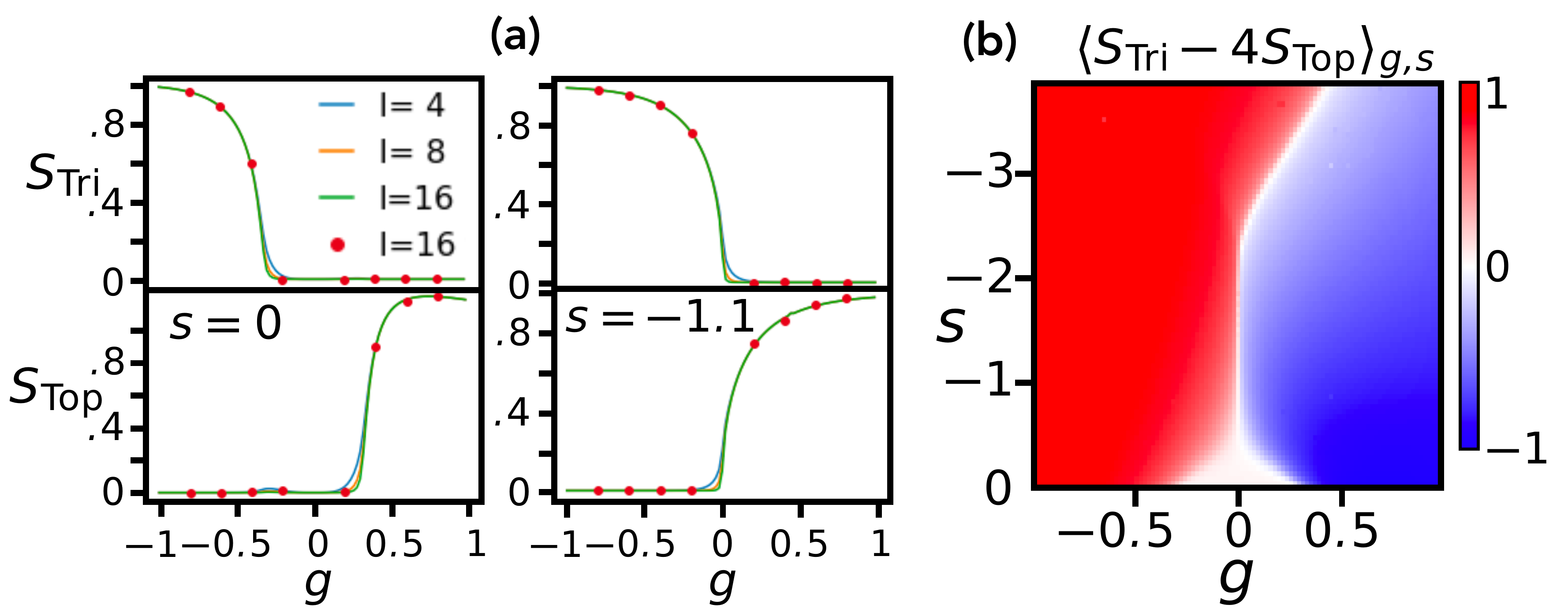}
    \caption{
        (Colour online) (a) Trivial string-order parameter $\Str$ (top) and non-trivial string-order parameter $\Sntr$ (bottom) at $\phi=1/2$. The left column is for $s=0$ and the right column for $s=-1.1$. Lines correspond to iMPS results for string lengths $\ell=4,8,16$ (blue, orange, green), while symbols a single sampled trajectory ($N=128$, $t=10^3$) at same conditions. (b) Phase diagram in terms of string-order parameters, 
        $\langle \Str - 4 \Sntr \rangle_{g,s}$ (for $\ell = 16$).
        } 
    \label{fig2}
\end{figure}

\smallskip 

\noindent 
{\bf\em LD phase diagram and SPT phases.--} The SCGF $\theta(g,s)$ for joint $\K^{(2)}$ and $\R$, where $g$ and $s$ are their corresponding conjugate (or counting) fields, is given by the largest eigenvalue of the tilted generator,
\begin{align}
    \WW_{g,s}
    &=
    \sum_j e^{g (-1)^j} \JJ_j - (1+s) \RR
    \nonumber \\    
    &= 
    \frac{1}{2} \sum_j \left[ 
        e^{g(-1)^j} \left( X_j X_{j+1} + Y_j Y_{j+1} \right) 
    \right. 
    \label{Wgs}
    \\
    & \;\;\;\;\;\;\;\;\;\;\;
    \left. 
        \phantom{e^{g(-1)^j}}
        + (1+s) \left( Z_j Z_{j+1} - 1 \right) 
    \right] ,
    \nonumber
\end{align}
which up to a sign is the Hamiltonian of an XXZ model with the terms in the kinetic energy staggered according to the counting factors $e^{g(-1)^j}$ 
\footnote{Cf.\ the bond-alternating XXZ model, see e.g.\ Refs.~\cite{Qiang2013, Tzeng2016}. Note that in contrast to these works, in our case there is no staggering in the diagonal terms, see \er{Wgs}.
}. The symmetries of this model include translation by two lattice sites, $U(1) \times {\mathbb Z}_2$ spin rotation symmetry, and time-reversal symmetry.
Since \er{Wgs} is Hermitian and short-ranged 
we can compute its largest eigenvalue $\theta(g,s)$ and its eigenvector $|R_{g,s}\rangle$ accurately using the density-matrix renormalization group method (DMRG) \cite{White_1992} by minimizing $-\WW_{g,s}$ as if it were a Hamiltonian. 
We work directly in the thermodynamic limit, $N \to \infty$, by approximating $|R_{g,s}\rangle$ as an infinite MPS (iMPS) 
with translationally invariant modulo two tensors, $B_n^{(1)}$ and $B_n^{(2)}$, to account for the staggering in $\WW_{g,s}$.

In Fig.~1(a), we map out the LD phase diagram in terms of $g$ and $s$, using infinite-system DMRG simulations of $|R_{g,s}\rangle$ (calculated using the TeNPy package \cite{Hauschild2018}). The case of $g = 0$ was studied before \cite{Appert-Rolland2008, Lecomte2012, Jack2015}, and we recover the first order transition at $s = 0$ between an inactive phase (IP) at $s > 0$ where particles are clustered, and an active critical phase (CP) for $s < 0$ with ``hyperuniform'' structure (a Luttinger liquid phase in the language of the XXZ model \cite{Sachdev2011}). As we consider $\R$ rather than the total number of hops as a measure of dynamical activity
\footnote{
    The tilted generator for the total number of configuration changes (which we call $\K$) is $\tilde{\WW}_s = e^{-s} \JJ - \RR$. It is directly related to that for $\R$, $\tilde{\WW}_s = e^{-s} \WW_{0,e^s-1}$, and so are the corresponding SCGFs \cite{Garrahan2009}. When biasing w.r.t.\ $\K$ the most active limit is given by $\tilde{\WW}_{-\infty} \propto \WW_{0,-1}$. When using $\R$ we can extend all the way to $\WW_{0,-\infty}$, allowing us to explore much further into the active phase.
}, 
we find another transition, which is of Kosterlitz-Thouless type,  deep in the active regime to a phase of maximal activity (MaxA) with antiferromagnetic order. 

For $g \neq 0$ we find two gapped phases, which have short-range correlations, cf.\ Fig.~2, and, in contrast to IP and MaxA, they are not symmetry-broken phases. As we explain below, they correspond two distinct SPT phases. Depending on the value of $s$ there is either a line of critical points between the two phases at $g = 0$ for $s < s_{\rm L}=1/\sqrt{2}-1$, or they are separated by a critical phase around $g = 0$ for $s > s_{\rm L}$. This follows from the fact that $|g| \gtrsim 0$ perturbations are relevant (in the RG sense) for $s < s_{\rm L}$ and irrelevant for $s > s_{\rm L}$. The  numerical value of $s_{\rm L}=1/\sqrt{2}-1$ is obtained from bosonization of the XXZ chain (see e.g.\ Refs.~\cite{Zamolodchikov_1995,Takayoshi_2010} and references therein). Figure 1(a) is for a bond dimension of $\chi=64$, which is sufficient to get a good approximation of $|R_{g,s}\rangle$. 

Figure 2(a) shows the average hop imbalance $\kappa = \lim_{t,N \to \infty}\langle \K^{(L)} \rangle_{g,s}/(N t)$ for two cuts in the phase diagram at fixed $s$ for two values of $\chi$. We notice that these curves are smooth as they cross $s=0$. Figure 2(b) plots the average hop imbalance $\kappa$ on the same phase diagram of Fig.~1(a), while Fig.\ 2(c) does so for the average escape rate, $\rho = \lim_{t,N \to \infty}\langle \R \rangle_{g,s}/(N t)$ 
\footnote{
    The average number of staggered jumps in the trajectory ensemble tilted by $g$ and $s$, per unit time and in the long time limit, is 
        $\lim_{t \to \infty} t^{-1} \langle \K^{(L)} \rangle_{g,s} 
            = -\partial_g \theta(g,s)
            = \sum_j f_j^{(L)} e^{-g f_j^{(L)}} \langle L_{g,s} | \JJ_j | R_{g,s} \rangle$, 
    where $\langle L_{g,s} |$ is the left leading eigenvector or $\WW_{g,s}$ (in our case $\langle L_{g,s} | = 2^N | R_{g,s} \rangle^\dagger$). 
    Similarly, for the average time-integrated escape rate, we have
    $\lim_{t \to \infty} t^{-1} \langle \R \rangle_{g,s} 
    = -\partial_s \theta(g,s)
    = \langle L_{g,s} | \RR | R_{g,s} \rangle$. 
    The iMPS provides an accurate approximation of these quantities per unit size in the large size limit.
}. 
While these average dynamical observables show the discontinuous change along the $s$ direction, they are smooth along the $g$ direction. The reason is that $\kappa$ and $\rho$ correspond to averages of local operators in the leading eigenstate of \er{Wgs} \cite{Note4} and cannot  distinguish between the SPT phases which do not break any symmetries of $\WW_{g,s}$.

\smallskip

\noindent 
{\bf\em String order parameters.--} To classify SPT phases, one needs to study instead averages of non-local observables, specifically string order parameters defined as follows. For filling $\phi = 1/2$ we consider cells of two contiguous sites, cf.\ Fig.~1(b), labelled by $k$, and define the total $Z_{k}^{(2)} = Z_{Lk} + Z_{Lk+1}$. The  string operator, which characterizes SPT phases protected by the ${\mathbb Z}_2\times{\mathbb Z}_2$ spin rotation symmetry \cite{denNijs89}, of length $\ell$ starting at cell $k$ is given by
\begin{align}
    S_{A}^{(k)}(\ell) = A_{k} e^{i \pi Z^{(2)}_k} e^{i \pi Z^{(2)}_{k+1}} 
            \cdots e^{i \pi Z^{(2)}_{k+\ell-1}} A_{k+\ell}
    \label{SO}
\end{align}
with either $A_k = 1$ (we call the corresponding string operator {\em trivial}, or $\Str$), or $A_k = Z^{(2)}_k$ (we call the corresponding string operator {\em non-trivial}, or $\Sntr$). Figure 3(a) shows the average $\langle \Str \rangle_{g,s}$ and $\langle \Sntr \rangle_{g,s}$ as function of $g$. We see that $\Str$ is non-zero in the ``Tri'' phase and $\Sntr$ is non-zero in the ``Top'' phase, with a change that tends towards becoming singular with increasing string length $\ell$. When the phase diagram, Fig.~3(b), is shown in terms of the string order parameters (plotted as $\langle \Str - 4 \Sntr \rangle_{g,s}$ to span $[-1,1]$) the transition between the ``Triv'' and ``Top'' phases becomes apparent (note that the AFM state is still symmetric under $\pi$-rotation around the $z$-axis and therefore has $\Str=1$).

\begin{figure*}[t]
    \centering
    \includegraphics[width=\textwidth]{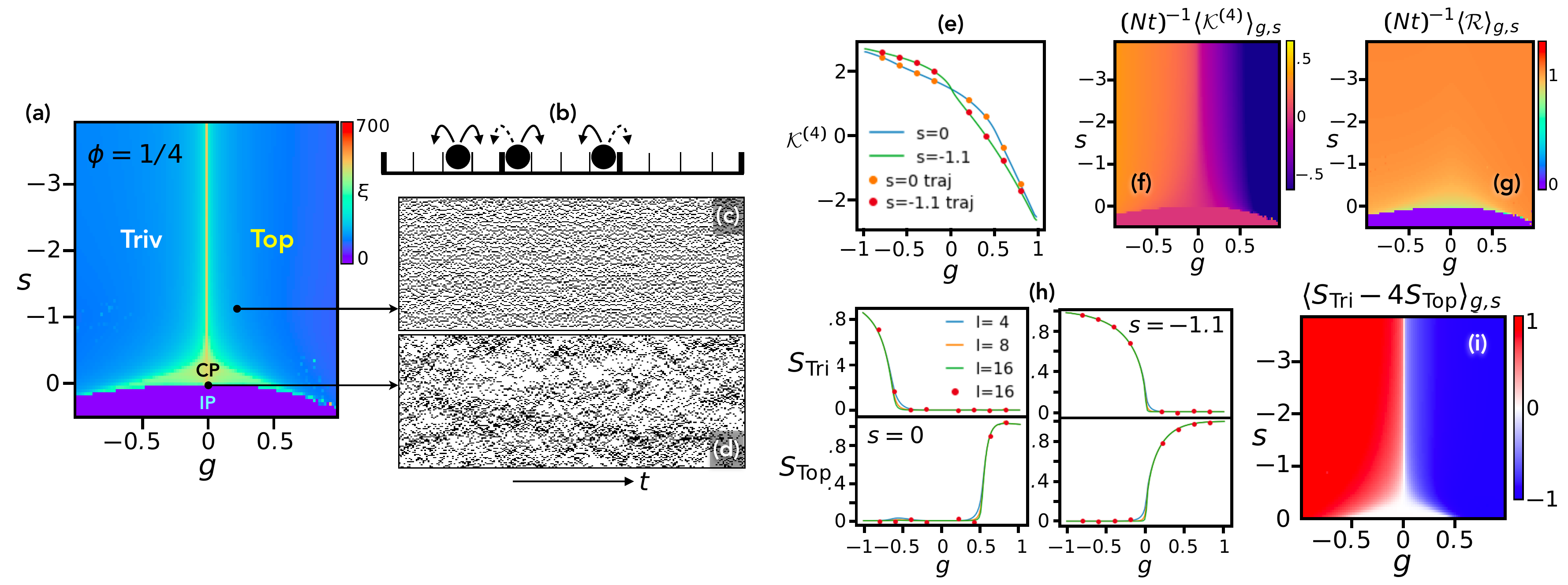}
    \caption{
        (Colour online) Same as Figs.~1-3 but for quarter-filling, $\phi = 1/4$.
        (a) LD phase diagram. As there is no $g \to - g$ symmetry there isno AFM phase. 
        (b) For $\K^{(4)}$ particle jumps across bonds that are $0$ mod $4$ (dashed arrows) count $-1$, while other jumps (full arrows) count $+1/3$. (c) Rare event trajectory from ``Top'' ($g=0.2, s=1.1$) sampled with (approximately) optimal dynamics, $N=128, t_{\rm max}=10^4$. (d) Typical trajectory at for comparison. 
        (e) Average staggered number of jumps $\kappa$ as a function of $g$ for $s=0$ (blue) and $s=-1.1$ (green) for half-filling $\phi=1/2$ (from iMPS with $\chi=128$). Symbols are from optimally sampled trajectories ($N=128$, $t=10^3$). (f,g) Phase diagrams in terms of average $\kappa$ and $\rho$.
        (h) String-order parameter for $s=0$ (left) and $s=-1.1$ (right). Symbols are from optimally sampled trajectories ($N=128$, $t=10^3$). (i) Phase diagram in terms of string-order parameters.
    }
    \label{fig3}
\end{figure*}


\smallskip

\noindent 
{\bf\em Doob transform and optimal sampling.--} 
We can also see how the SPT phases manifest at the level of trajectories. Sampling rare trajectories corresponding to $g \neq 0$ (and/or $s \neq 0$) is exponentially expensive in both time and system size. However, given that the iMPS provides a very good approximation of the leading eigenvectors 
$|R_{g,s}\rangle$ and $\langle L_{g,s}|$, we can construct the dynamics that optimally samples rare trajectories at $g,s$ via the (long-time) Doob transform
\cite{Jack2010, Chetrite2015, Garrahan2016, Carollo2018}
\begin{align}
    \tilde{\WW}(g,s) 
    &=
    \sum_j e^{g (-1)^j} \LL_{g,s} \, \JJ_j \, \LL^{-1}_{g,s}
    - (1 + s) \RR - \theta(g,s)
    \label{Doob}
 \end{align}
where $\LL_{g,s}$ is a diagonal matrix, $[\LL_{g,s}]_{xy} = \delta_{xy} L_{g,s}(x)$, of the components of $\langle L_{g,s}| = \sum_x L_{g,s}(x) \langle x |$. The generator above is stochastic and its corresponding trajectories can be sampled directly (e.g.\ by continuous-time Monte Carlo). In practice, we write the components of $\LL$ in terms of the iMPS that maximises \er{Wgs}, $L_{g,s}(x = n_{1:N}) = \Tr B^{(1)}_{n_1} B^{(2)}_{n_2} \cdots B^{(1)}_{n_{N-1}} B^{(2)}_{n_N}$ (for system size $N$ even). The accuracy of the iMPS means that this is an excellent approximation of the exact Doob transition rates (see also \cite{Causer2021}).
Figure 1(c) shows a trajectory representative of the ``Top'' phase for $\phi = 1/2$. 
These rare trajectories can be generated on demand using continuous-time Monte Carlo with rates from $\tilde{\WW}(g,s)$. The trajectory in Fig.~1(c) looks very different from a typical trajectory, cf.\ Fig.~1(d). The quality of this quasi-optimal sampling can be seen in Fig.~2(a), where the values for the staggered number of jumps from a single long trajectory at each state point coincide with those from the iMPS.

Interestingly, while the topological character of an SPT phase is a property of the eigenvector $|R_{g,s}\rangle$ that characterises the whole dynamical phase, the string order is present already at the level of a single trajectory at the corresponding conditions: Fig.~3(a) shows that the values of $\Str$ and $\Sntr$ averaged over one long atypical trajectory at $g \neq$, sampled efficiently using the Doob dynamics, coincide with the DMRG results.

\smallskip

\noindent 
{\bf\em Generalisation to other filling fractions.--} We can generalize the results above for particle densities different than half-filling. For example, for filling fraction $\phi = 1/L$, the appropriate observable conjugate to $g$ is given by $\K^{(L)} = \sum_j f_j K_j$, with $f_j^{(L)}=-1$ if $j=\modL{0}$ and $f_j^{(L)} = 1/(L-1)$, which counts the difference in activity within and across cells of size $L$, see Fig.~1(b).
The tilted generator is then 
$\WW_{g,s} = \sum_j e^{-g f_j^{(L)}} \JJ_j - (1+s) \RR$, and we approximate the eigenstate $|R_{g,s}\rangle$ by an iMPS with tensors $B_n^{(1)}, \cdots, B_n^{(L)}$.

Figure 4 shows the case of quarter-filling, $\phi = 1/4$. The LD phase diagram is similar to that of the half-filling case, with two notable differences: (i) there is no $g \to -g$ symmetry; (ii) the MaxA antiferromagnetic phase is absent, again due to symmetry arguments. (Using the bosonization method of e.g.\ Ref.~\cite{Takayoshi_2010} the value of $s_{\rm L}$ that delimits CP could be calculated.)
As for half-filling, the trivial and topological phases can be distinguished via string order parameters: If we group sites into unit cells of size $L$ the total spin in cell $k$ is $Z_{k}^{(L)} = Z_{Lk} + Z_{Lk+1} + \cdots + Z_{L(k+1)-1}$. The string operators for $\phi=1/L$ are then defined as in \er{SO} replacing $Z_{k}^{(2)}$ by $Z_{k}^{(L)}$ and using $A_k = Z^{(L)}_k + \frac{1}{2} L - 1$ for the endpoints of $\Sntr$. Figures 4(h,i) show the average string order for $\phi = 1/4$. As in the case of half-filling, from the iMPS we can construct the optimal sampling dynamics \er{Doob}: Fig.~4(c) shows an atypical trajectory from the SPT  phase in the $\phi=1/4$ case, and 
Figs.~4(e,h) show that values of average dynamical observables and string order parameters can be obtained from sampling long-time rare trajectories also in the quarter-filling case.

\smallskip

\noindent 
{\bf\em Outlook.--} An interesting question is whether 
other stochastic models can also have topological dynamical phases, like those above that arise in the symmetric SEP in atypical trajectories with spatially modulated patterns of activity. Here we exploited the Hermiticity of the SSEP to map out these dynamical phases and the transitions between them by means of standard
MPS techniques. Interestingly, properties of the SPT phases (for example their string order) is manifest at the level of individual rare trajectories, which makes them in principle directly observable. We can anticipate that for one version of the ASEP the symmetric trivial and topological phases of the SSEP should also be present: with open boundaries and with no injection/ejection of particles, the ASEP can be mapped to a (tilted) SSEP (see e.g.\ \cite{De-Gier2006})
(with extra boundary terms which should not matter in the large size limit); this mapping extends to the tilted generator, cf.\ \er{Wgs}, and the LD phase diagram of this ASEP should then be like that of the SSEP, Figs.~1(a) and 4(a), but shifted vertically. 
A more difficult step will be to establish similar results in genuinely driven systems.

\smallskip

\noindent 
{\bf\em Acknowledgements.--}
We thank Pablo Sala and Ruben Verresen for fruitful discussions. JPG acknowledges financial support from EPSRC Grant no.\ EP/R04421X/1 and the Leverhulme Trust Grant No.\ RPG-2018-181. We acknowledge access to the University of Nottingham Augusta HPC service. This work was supported by the European Research Council (ERC) under the European Union’s Horizon 2020 research and innovation program (grant agreement No. 771537). F.P. acknowledges the support of the Deutsche Forschungsgemeinschaft (DFG, German Research Foundation) under Germany’s Excellence Strategy EXC-2111-390814868. The research is part of the Munich Quantum Valley, which is supported by the Bavarian state government with funds from the Hightech Agenda Bayern Plus.

\bibliography{SPT_SEP}
\bibliographystyle{apsrev4-1}

\end{document}